support@word2tex.com

\documentclass [12pt]{article}

\begin{document}
\begin{center}
\textbf{\huge Generalized uncertainty principle and correction
value to the black hole entropy}

\end{center}
\begin{center}
Zhao Hai-Xia$^1$, Li Huai-Fan$^2$, Hu Shuang-Qi$^{1)}$ and Zhao
Ren$^{1,2)}$\footnote{Corresponding address: Department of
Physics, Shanxi Datong
University, Datong 037009 P.R.China;\\
E-mail address: zhaoren2969@yahoo.com.cn }

1) School of Chemical Engineering and Environment, North
Universityof China, Taiyuan 030051 \\
2) Department of Physics, Shanxi Datong University, Datong 037009

\end{center}

\begin{center}
\textbf{Abstract}
\end{center}
Recently, there has been much attention devoted to resolving the
quantum corrections to the Bekenstein-Hawking  entropy of the
black hole. In particular, many researchers have expressed a
vested interest in the coefficient of the logarithmic term of the
black hole entropy correction term. In this paper, we calculate
the correction value of the black hole entropy by utilizing the
generalized uncertainty principle and obtain the correction term
caused by the generalized uncertainty principle. Because in our
calculation we think that the Bekenstein-Hawking area theorem is
still valid after considering the generalized uncertainty
principle, we derive that the coefficient of the logarithmic term
of the black hole entropy correction term is negative. This result
is different from the known result at present. Our method is valid
not only for single horizon spacetime but also for double horizons
spacetime. In the whole process, the physics idea is clear and
calculation is simple. It offers a new way for studying the
condition that Bekenstein-Hawking area theorem is valid.\vspace{1.cm}
\vspace{1.cm}\\
\textbf{\Large 1. Introduction}

One of the most remarkable achievements in gravitational physics
was the realization that black holes have temperature and
entropy$^{ [1 - 3]}$. There is a growing interest in the black
hole entropy. Because entropy has statistical physics meaning in
the thermodynamic system, it is related to the number of
microstates of the system. However, in Einstein general relativity
theory, the black hole entropy is a pure geometry quantity. If we
compare the black hole with the thermodynamic system, we will find
an important difference. A black hole is a vacancy with strong
gravitation. But the thermodynamic system is composed of atoms and
molecules. Based on the microstructure of thermodynamic systems,
we can explain thermodynamic property by statistic mechanics of
its microcosmic elements. Whether the black hole has interior
freedom degree corresponding the black hole entropy?$^{[4]}$ let
us suppose that the Bekenstein-Hawking entropy can be attributed a
definite statistical meaning. Then how might one go about
identifying these microstates and, even more optimistically,
counting them?$^{[5]}$ This is a key problem to study the black
hole entropy.

Recently, string theory and loop quantum gravity have both had
success at statistically explaining the entropy-area
``law''$^{[5]}$. However, who might actually prefer if there was
only one fundamental theory? It is expected to choose it by
quantum correction term of the black hole entropy. Therefore,
studying the black hole entropy correction value becomes the focus
of attention. Many ways of discussing the black hole entropy
correction value have emerged$^{ [5 - 12]}$. But the exact value
of coefficient of the logarithmic term in the black hole entropy
correction term is not known. Since we discuss the black hole
entropy, we need study the quantum effect of the black hole. When
we discuss radiation particles or absorption ones, we should
consider the uncertainty principle. However, as gravity is turned
on, the ``conventional'' Heisenberg relation is no longer
completely satisfactory. The generalized uncertainty principle
will replace it. In this paper, we discuss the black hole entropy
correction value by the generalized uncertainty principle. There
is no restriction to spacetimes in the method given by us. So our
result has general meaning. We will discuss three kinds of
representative spacetimes. The paper is organized as follows.
Section 2 analyses Schwarzschild spacetime. Section 3 discusses
Carfinkle-Horowitz-Strominger. Section 4 describes
Reissner-Nordstr$\ddot {o}$m spacetime with double horizons. We
take the simple function form of temperature ($c= \hbar = G = K_B
= 1)$.\vspace{1.cm}\\
\textbf{\Large 2. Schwarzschild Black hole}

The linear element of Schwarzschild black hole spacetime:

\begin{equation}
\label{eq1} ds^2 = - \left( {1 - \frac{2M}{r}} \right)dt^2 +
\left( {1 - \frac{2M}{r}} \right)^{ - 1}dr^2 + r^2d\Omega _2^2 .
\end{equation}
Hawking radiation temperature $T$, horizon area $A$ and entropy
$S$ are respectively

\begin{equation}
\label{eq2} T = \frac{1}{4\pi r_H } = \frac{1}{8\pi M}, \quad A =
4\pi r_H^2 = 16\pi M^2, \quad S = \pi r_H^2 = 4\pi M^2,
\end{equation}

\noindent where $r_H = 2M$ is the location of the black hole
horizon.

Now for a black hole absorbing (radiating) a particle of energy
$dM$, the increase (decrease) in the horizon area can be expressed
as

\begin{equation}
\label{eq3} dA = 8\pi r_H dr_H = 32\pi MdM.
\end{equation}
Because the discussed black hole radiation is a quantum effect,
the particle of energy $dM$ should satisfy Heisenberg uncertainty
relation.

\begin{equation}
\label{eq4} \Delta x_i \Delta p_j \ge \delta _{ij} .
\end{equation}
In gravity field Heisenberg uncertainty relation should be
replaced by the generalized uncertainty principle $^{[13-15]}$:

\begin{equation}
\label{eq5} \Delta x_i \ge \frac{\hbar }{\Delta p_i } + \alpha
^2l_{pl}^2 \frac{\Delta p_i }{\hbar },
\end{equation}

\noindent where $l_{pl} = \left( {\textstyle{{\hbar G_d } \over
{C^3}}} \right)^{1 / 2}$ is Planck length, $\alpha $ is a
constant. From (\ref{eq5}), we obtain

\begin{equation}
\label{eq6} \frac{\Delta x_i }{2\alpha ^2l_{pl}^2 }\left[ {1 -
\sqrt {1 - \frac{4\alpha ^2l_{pl}^2 }{\Delta x_i^2 }} } \right]
\le \frac{\Delta p_i }{\hbar } \le \frac{\Delta x_i }{2\alpha
^2l_{pl}^2 }\left[ {1 + \sqrt {1 - \frac{4\alpha ^2l_{pl}^2
}{\Delta x_i^2 }} } \right].
\end{equation}
At $\alpha = 0$, we express (\ref{eq6}) by Taylor series and
derive

\begin{equation}
\label{eq7} \Delta p_i \ge \frac{1}{\Delta x_i }\left[ {1 + \left(
{\frac{\alpha ^2l_{pl}^2 }{(\Delta x_i )^2}} \right) + 2\left(
{\frac{\alpha ^2l_{pl}^2 }{(\Delta x_i )^2}} \right)^2 + \cdots }
\right].
\end{equation}
From (\ref{eq3}) and (\ref{eq4}), the change of the area of the
black hole horizon can be written as follows:

\begin{equation}
\label{eq8} dA = 8\pi r_H dr_H = 32\pi Mdp = 32\pi
M\frac{1}{\Delta x}.
\end{equation}
According to the generalized uncertainty principle (\ref{eq7}) and
(\ref{eq3}), the change of the area of the black hole horizon can
be rewritten as follows:

\begin{equation}
\label{eq9} dA_G = 8\pi r_H dr_H = 32\pi Mdp = 32\pi M
\frac{1}{\Delta x}\left[ {1 + \left( {\frac{\alpha ^2l_{pl}^2
}{(\Delta x)^2}} \right) + 2\left( {\frac{\alpha ^2l_{pl}^2
}{(\Delta x)^2}} \right)^2 + \cdots } \right].
\end{equation}
From (\ref{eq8}) and (\ref{eq9}), we derive

\begin{equation}
\label{eq10} dA_G = \left[ {1 + \left( {\frac{\alpha ^2l_{pl}^2
}{(\Delta x)^2}} \right) + 2\left( {\frac{\alpha ^2l_{pl}^2
}{(\Delta x)^2}} \right)^2 + \cdots } \right]dA.
\end{equation}
According the view of Ref.[5,8], we take

\begin{equation}
\label{eq11} \Delta x = 2r_H = 2\sqrt {\frac{A}{4\pi }} .
\end{equation}
Substituting (\ref{eq11}) into (\ref{eq10}) and integrating, we
derive

\begin{equation}
\label{eq12} A_G = A + \alpha ^2l_{pl}^2 \pi \ln A
 - 2(\alpha ^2l_{pl}^2 \pi )^2\frac{1}{A} - \cdots .
\end{equation}
Based on Bekenstein-Hawking area law, we take $S = A / 4$.
Therefore, we can derive the expression of entropy after
considering the generalized uncertainty principle. That is, the
correction to entropy is given by

\begin{equation}
\label{eq13} S_G = S + \alpha ^2l_{pl}^2 \pi \ln S - (\alpha
^2l_{pl}^2 \pi )^2\frac{1}{2S} - \cdots + C.
\end{equation}
Where $S$ is Bekenstein-Hawking entropy, $C$ is an arbitrary
constant. In the calculation, we can plus or minus an arbitrary
constant $C$. From (\ref{eq13}), we can calculate an arbitrary
term of correction to entropy and obtain that the coefficient of
the logarithmic correction term is positive. This result is
different from that of Ref.[5].\vspace{1.cm}\\
\textbf{\Large 3. Garfinkle-Horowitz-Strominger dilaton black
hole}

The linear element in Garfinkle-Horowitz-Strominger dilaton black
hole spacetime is$^{[16,17]}$:

\begin{equation}
\label{eq14} ds^2 = - \left( {1 - \frac{2M}{r}} \right)dt^2 +
\left( {1 - \frac{2M}{r}} \right)^{ - 1}dr^2 + r(r - 2a)d\Omega
_2^2 ,
\end{equation}

\noindent where $a = Q^2 / 2M$, $Q$ is the electric charge of the
black hole. The horizon area $A$ and entropy $S$ of the black hole
are respectively

\begin{equation}
\label{eq15} A = 4\pi r_H (r_H - 2a) = 16\pi M(M - a), \quad S =
4\pi M(M - a).
\end{equation}
Under the case that $a$ is invariable, we obtain

\begin{equation}
\label{eq16} dA = 16\pi (2M - a)dM.
\end{equation}
Considering the generalized uncertainty principle, we have

\begin{equation}
\label{eq17} dA_G = \left[ {1 + \left( {\frac{\alpha ^2l_{pl}^2
}{(\Delta x)^2}} \right) + 2\left( {\frac{\alpha ^2l_{pl}^2
}{(\Delta x)^2}} \right)^2 + \cdots } \right]dA.
\end{equation}
Let $\Delta x = 2r_H $, (\ref{eq17}) can be rewritten as

\[
dA_G = \left[ {1 + \frac{\alpha ^2l_{pl}^2 \pi }{A} -
\frac{8\alpha ^2l_{pl}^2 \pi ^2ar_H }{A^2}} \right. \left. { +
2\frac{(\alpha ^2l_{pl}^2 )^2\pi ^2}{A^2} - 2\frac{16(\alpha
^2l_{pl}^2 )^2a^2r_H^2 \pi ^4}{A^4} + \cdots } \right]
\]

\[
 \approx \left[ {1 + \frac{\alpha ^2l_{pl}^2 \pi }{A} - \frac{8\alpha
^2l_{pl}^2 \pi ^2a^2}{A^2} - \frac{4\alpha ^2l_{pl}^2 \pi ^{3 /
2}a}{A^{3 / 2}}} \right. \left. { + 2\frac{(\alpha ^2l_{pl}^2
)^2\pi ^2}{A^2} - 2\frac{4(\alpha ^2l_{pl}^2 )^2a^2\pi ^3}{A^3} +
\cdots } \right]dA.
\]
Thus

\[
A_G = A + \alpha ^2l_{pl}^2 \pi \ln A
 + \frac{8\alpha ^2l_{pl}^2 \pi ^2a^2}{A}
 + \frac{8\alpha ^2l_{pl}^2 \pi ^{3 / 2}a}{A^{1 / 2}}
\]

\begin{equation}
\label{eq18}
 - 2\frac{(\alpha ^2l_{pl}^2 \pi )^2}{A} + \frac{4(\alpha ^2l_{pl}^2
\pi a)^2\pi }{A^2} + \cdots .
\end{equation}
In calculation, we have used the relation $a < < r_H $. According
to Bekenstein-Hawking area law, we take $S = A / 4$. Therefore, we
can derive the expression of entropy after considering the
generalized uncertainty principle. That is, the correction to
entropy is given by

\[
S_G = S + \alpha ^2l_{pl}^2 \pi \ln S
 + \frac{8\alpha ^2l_{pl}^2 \pi ^2a^2}{S}
 + \frac{8\alpha ^2l_{pl}^2 \pi ^{3 / 2}a}{S^{1 / 2}}
\]

\begin{equation}
\label{eq19}
 - 2\frac{(\alpha ^2l_{pl}^2 \pi )^2}{S} + \frac{4(\alpha ^2l_{pl}^2
\pi a)^2\pi }{S^2} + C \cdots .
\end{equation}
Where $C$ is an arbitrary constant.\vspace{1.cm}\\
\textbf{\Large 3. Reissner-Nordstr$\ddot {o}$m black hole}

The linear element in Reissner-Nordstr$\ddot {o}$m black hole
spacetime is given by:

\begin{equation}
\label{eq20} ds^2 = - \left( {1 - \frac{2M}{r} + \frac{Q^2}{r^2}}
\right)dt^2 + \left( {1 - \frac{2M}{r} + \frac{Q^2}{r^2}}
\right)^{ - 1}dr^2 + r^2d\Omega _2^2 ,
\end{equation}
where $r_\pm = M\pm \sqrt {M^2 - Q^2} $ are the locations of outer
and inner horizons respectively. $Q$ is the electric charge of the
black hole. The outer horizon area $A$ and entropy $S$ of the
black hole are respectively

\begin{equation}
\label{eq21} A = 4\pi r_ + ^2 , \quad S = \pi r_ + ^2 .
\end{equation}
When $Q$ is invariable,

\begin{equation}
\label{eq22} dA = 16\pi \frac{r_ + ^2 }{r_ + - r_ - }dM.
\end{equation}
Considering the generalized uncertainty principle, we derive

\begin{equation}
\label{eq23} dA_G = \left[ {1 + \left( {\frac{\alpha ^2l_{pl}^2
}{(\Delta x)^2}} \right) + 2\left( {\frac{\alpha ^2l_{pl}^2
}{(\Delta x)^2}} \right)^2 + \cdots } \right]dA.
\end{equation}
Let $\Delta x = 2(r_ + - r_ - )$, (\ref{eq23}) can be rewritten as

\begin{equation}
\label{eq24} dA_G = \left[ {1 + (\alpha ^2l_{pl}^2 )\left(
{\frac{\pi }{A} + \frac{8\pi Q^2}{A^2} - \frac{16\pi ^2Q^4}{A^3}}
\right)} \right. + \left. {2(\alpha ^2l_{pl}^2 )^2\left(
{\frac{\pi }{A} + \frac{8\pi Q^2}{A^2} - \frac{16\pi ^2Q^4}{A^3}}
\right)^2 + \cdots } \right]dA.
\end{equation}
Thus

\[
A_G = A + \alpha ^2l_{pl}^2 \pi \ln A
 - (\alpha ^2l_{pl}^2 )\left( {\frac{8\pi Q^2}{A} - \frac{8\pi
^2Q^4}{A^2}} \right)
\]

\begin{equation}
\label{eq25}
 - 2(\alpha ^2l_{pl}^2 \pi )^2\left( {\frac{1}{A} + \frac{8Q^2}{A^2} +
\frac{64Q^4}{3A^3} - \frac{32\pi Q^4}{3A^3}} \right) + \cdots .
\end{equation}
In calculation, we have used the relation $Q < < r_H $. According
to Bekenstein-Hawking area law, we take $S = A / 4$. Therefore, we
can derive the expression of entropy after considering the
generalized uncertainty principle. That is, the correction to
entropy is given by

\[
S_G = S + \alpha ^2l_{pl}^2 \pi \ln S - (\alpha ^2l_{pl}^2 )\left(
{\frac{8\pi Q^2}{S} - \frac{8\pi ^2Q^4}{S^2}} \right)
\]

\begin{equation}
\label{eq26}
 - 2(\alpha ^2l_{pl}^2 \pi )^2\left( {\frac{1}{S} + \frac{8Q^2}{S^2} +
\frac{64Q^4}{3S^3} - \frac{32\pi Q^4}{3S^3}} \right) + C + \cdots
.
\end{equation}
Where $C$ is an arbitrary constant.\vspace{1.cm}\\
\textbf{\Large 4. Conclusion}

In summary, we have utilized the generalized uncertainty principle
to demonstrate an explicit form for the correction term of the
black hole entropy. From (\ref{eq13}), (\ref{eq18}) and
(\ref{eq26}), for different spacetimes, coefficients of the
logarithmic terms in the black hole entropy correction terms are
same and are positive. This result is different from that of
Ref.[5]. Although this paper and Ref.[5] both discuss the
corrections caused by the generalized uncertainty principle to the
black hole entropy. The difference between this paper and Ref.[5]
is as follows: In (\ref{eq13}) given by Ref.[5], the ratio between
the black hole entropy and the horizon area is not quarter and
there is a correction value; in our result the ratio between the
black hole entropy and the horizon area is always quarter without
reference to whether considering the generalized uncertainty
principle or not. That is, Bekenstein-Hawking area law after
considering the generalized uncertainty principle is always valid.
Thus, we derive that coefficient of the logarithmic term in the
black hole entropy correction term is positive. This is different
with the logarithmic correction being negative.

Based on the above analysis, we calculate the correction term of
the black hole entropy under the condition that Bekenstein-Hawking
area law after considering the generalized uncertainty principle
is valid and there are no others assumptions. So our calculation
is reliable. Our method is valid not only for single horizon
spacetime but also for unextreme spacetimes with outer and inner
horizons. It offers a new way for studying the entropy correction
of the complicated spacetime.

If we can obtain the exact value of the coefficient of the
logarithmic term in the black hole entropy correction term by
other method, we can not only determine the uncertainty number
$\alpha $ in the generalized uncertainty principle but also obtain
the condition that Bekenstein-Hawking area law is valid.
Therefore, our results provide a new subject for studying the
condition that Bekenstein-Hawking area law is valid.\\
\textbf{ACKNOWLEDGMENT}

This project was supported by the National Natural Science
Foundation of China under Grant No. 10374075 and the Shanxi
Natural Science Foundation of China (No 2006011012).

\textbf{References}

[1] J. D. Bekenstein,  Phys. Rev.  \textbf{D7}, 2333(1973)

[2] J. D. Bekenstein,   Phys. Rev. \textbf{D9}, 3292(1974)

[3] S. W. Hawking,   Nature \textbf{248}, 30(1974); Commun. Math.
Phys. \textbf{43}, 199(1975)

[4] Y. J. Wang,  Black Hole Physics (Changsha: Hunan Normal
University Press), p263(2000) (in Chinese)

[5] A. J. M.  Medved and E. C. Vagenas,  Phys. Rev.  \textbf{D70
}, 124021(2004)

[6] A. Chatterjee and P. Majumdar,  Phys. Rev. Lett. \textbf{92},
141301(2004)

[7] R. K. Kaul and P. Majumdar,   Phys. Rev. Lett. \textbf{84},
5255(2000)

[8] G. A. Camellia, M. Arzano  and A. Procaccini,   Phys. Rev.
\textbf{D70}, 107501(2004)

[9] A. Chatterjee  and P. Majumdar,   Phys. Rev. \textbf{D71},
024003(2005)

[10] Y. S. Myung,  Phys. Lett. \textbf{B579}, 205(2004)

[11] M. M. Akbar  and S. Das,   Class. Quant. Grav. \textbf{21},
1383(2004)

[12] S. Das,   Class. Quant. Grav. \textbf{19}, 2355(2002)

[13] L. N.Chang, D. Minic, N. Okaruma and  T. Takeuchi, Phys. Rev.
\textbf{D65}, 125028(2002)

[14] S. Q. Hu and R. Zhao,   Chinese Physics \textbf{14},
1477(2005)

[15] X. Li,   Phys. Lett. \textbf{B540}, 9(2002)

[16] D. Carfinkle, Horowitz and A. Strominger,  Phys. Rev.
D\textbf{43}, 3140(1991)

[17] S. Chen  and J. L. Jing,  Class. Quant. Grav. \textbf{22 },
533(2005)

\end{document}